\documentclass[10pt,journal,compsoc]{IEEEtran}
\usepackage{cite}
\usepackage{multicol}
\usepackage{graphicx}
\graphicspath{ {images/} }
\usepackage{amssymb}
\usepackage{algorithm,algorithmic}
\usepackage{mathtools}
\usepackage{mathptmx}
\usepackage{amsmath}
\usepackage{amsthm}
\usepackage{listings}
\usepackage{multirow}
\usepackage{amssymb}
\usepackage{amsthm}
\usepackage{subfigure}
\usepackage{caption}
\usepackage{url}
\usepackage{paralist}
\usepackage{xcolor}
\usepackage{svg}
\newtheorem{definition}{Definition}

\usepackage[font=small,labelfont=bf]{caption}

\begin{document}

\title{Sustainable Volunteer Engagement: Ensuring Potential Retention and Skill Diversity for Balanced Workforce Composition in Crowdsourcing Paradigm\vspace{-0.03in}}


\author{Riya~Samanta,~\IEEEmembership{Student Member,~IEEE,}
        Soumya~K.~Ghosh,~\IEEEmembership{Member,~IEEE,}
\thanks{Riya Samanta and Soumya K. Ghosh are with the Department
of Computer Science and Engineering, Indian Institute of Technology Kharagpur, India.\\
E-mail: \{riya.samanta, skg\}@cse.iitkgp.ac.in}
\vspace*{-0.1in}
}

\IEEEtitleabstractindextext{
\begin{abstract}
Crowdsourcing (CS) faces the challenge of managing complex, skill-demanding tasks, which requires effective task assignment and retention strategies to sustain a balanced workforce. This challenge has become more significant in Volunteer Crowdsourcing Services (VCS). This study introduces Workforce Composition Balance (WCB), a novel framework designed to maintain workforce diversity in VCS by dynamically adjusting retention decisions. The WCB framework integrates the Volunteer Retention and Value Enhancement (VRAVE) algorithm with advanced skill-based task assignment methods. It ensures efficient remuneration policy for both assigned and unassigned potential volunteers by incorporating their potential levels, participation dividends, and satisfaction scores. Comparative analysis with three state-of-the-art baselines on real dataset shows that our WCB framework achieves 1.4 times better volunteer satisfaction and a 20\% higher task retention rate, with only a 12\% increase in remuneration. The effectiveness of the proposed WCB approach is to enhance the volunteer engagement and their long-term retention, thus making it suitable for functioning of social good applications where a potential and skilled volunteer workforce is crucial for sustainable community services.
\end{abstract}

}

\maketitle
\footnote{This work has been submitted to the IEEE for possible publication. Copyright may be transferred without notice, after which this version may no longer be accessible.}
\IEEEdisplaynontitleabstractindextext
\IEEEpeerreviewmaketitle

\IEEEraisesectionheading{\section{Introduction}\label{intro}}
Crowdsourcing (CS) has become integral to gathering and analyzing diverse data for various social applications like volunteering, delivery logistics, and participatory sensing \cite{bedogni2023towards}. These activities involve diverse crowd workers who contribute actively or passively, dealing with spatial or non-spatial data in various formats and qualities.

For any CS-based social project, collaboration among numerous participants is crucial, forming a many-to-one matching problem based on social computing and market matching principles. Task assignment is extensively studied to ensure the right candidates are matched with appropriate tasks, whether one-to-one, one-to-many, or many-to-many. \textit{However, for the smooth functioning of a CS system, is merely initiating task-candidate pairings, aggregating outputs, ensuring work quality, and distributing rewards sufficient?}

Research  shows that most of the crowd workers on online CS platforms are primarily short-term participants. This deficiency in sustained crowd engagement has led to an imbalance in workforce composition, hindering large market enterprises from achieving their full utility \cite{shi2022motivates}. Crowd workers often disengage due to waning motivation or switch platforms for better perks and visibility, highlighting the need for strategies to select and retain competent workers \cite{nawaz2019gig}. While initial participation is vital in CS platforms, sustained crowd engagement is crucial for their long-term success and the continuous delivery of high-quality services for social good \cite{leung2021crowd,wu2022understanding}.

Volunteer Crowdsourcing Services (VCS), such as VolunteerMatch \cite{volmatch} and SkillImpact \cite{skillimpact}, are platforms within the growing community of citizen science that empower individuals to engage in data collection, analysis, and social work initiatives \cite{sultan2022addressing}. In this paradigm, particularly when \textit{server-assigned task mode} \cite{Chen2019} is followed, designing an effective task assignment and retention strategy is essential to sustainably maintain the volunteer workforce. This can be achieved by valuing their contributions through reimbursement or recognition and by modeling their latent potential for future goals \cite{Kaur2022}. Moreover, in case of complex and skill-demanding heterogeneous tasks, assignment become even more challenging because in that case candidates need to be chosen as per task's explicit competency and skill requirements \cite{samanta2021swill,Cheng2016,Song2020,Hettiachchi2022}. Thus, during retention, a proper balance among newcomers and experienced ones needs to be evaluated without discouraging any dropouts. In general, with respect to skill-based VCS ventures \cite{Sama2212:Volunteer}, volunteers assigned to complex, heterogeneous and skill-oriented CS tasks, primarily participate for moral values and gaining experience. Thus, task assignment and retention strategies should focus on providing a positive impact on the participants' \textit{satisfaction} levels \cite{Hettiachchi2022}. Beyond remuneration support, these strategies should be driven by the experience gained and the skills developed by the participants.

\begin{figure*}[!t]
\centering
{\includegraphics[width=\textwidth]{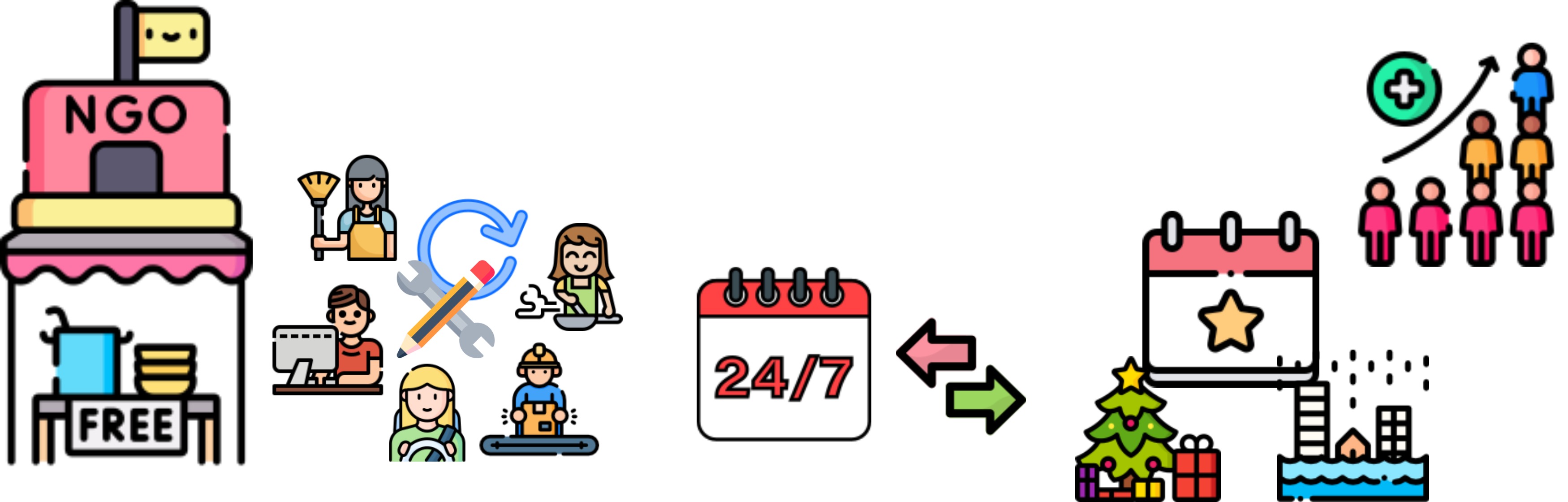}}
\caption{Food Bank: Skill-based Volunteer Crowdsourcing Service (VCS) Requiring Workforce Composition Balance for Regular Operations and Surge Events (Holidays, Natural Calamities, etc.) }
\label{food}
\vspace{-0.2in}
\end{figure*}

\vspace{-0.05in}
\subsection{Motivating Example}
\vspace{-0.05in}
Consider the example of a \textit{Food Bank} run by an NGO (see \textbf{Figure \ref{food}}), which serves as a standard model in the realm of skill-based VCS, relying on volunteers for both long-term and short-term initiatives \cite{Kaur2022,8423113}. Skill demand is a critical aspect in this context, as volunteers are tasked with various responsibilities such as sorting and packaging food, cooking, auditing, driving delivery vehicles, and managing logistics. Apart from addressing surge events like the distribution of food packages during specific high-demand periods (such as holidays, school breaks or natural calamities), the food bank requires a steady stream of volunteers to maintain long-term services. These services are supposed to operate on a regular schedule, such as opening for fixed hours on 2–3 designated days each week. While new volunteers are always welcomed, it is equally important to properly incentivize existing volunteers based on their latent potential and expected future contributions. 

\vspace{-0.05in}
\section{Limitations, Challenges and Our Research Questions}
\label{limit}
Existing research on workforce composition balance (see Section \ref{section2}, Related Work), has predominantly focused on micro-tasks. These studies often do not account for the true potential of crowd workers during retention decisions and tend to view the workforce as a homogeneous group. However, in practice, CS platforms consist of crowd workers with varying levels of potential based on their skills and experience. Additionally, the tasks handled are largely heterogeneous, with diverse requirements. Therefore, it is imperative to categorize \textit{newcomer} separately from \textit{existing crowd worker}. In this study, \textit{newcomers} refer to crowd workers who are registering for the first time on the CS platform, as well as those applying for a task in a particular domain for the first time. Recruiting such newcomers is typically time-consuming, requires stringent profile screening, and may entail additional costs, adding complexity to the task assignment phase.

This complexity is particularly pronounced in Volunteer Crowdsourcing Service (VCS) ventures, where greater flexibility is needed to leverage volunteers' contributions effectively. By addressing these challenges, our study aims to ensure a balanced and sustainable workforce composition on CS platforms.

Thus, in the context of VCS, the above discussion leads to the following \textbf{research questions}:

\begin{itemize}
    \item How can an existing task assignment policy be upgraded to ensure a balanced workforce composition among participants in the context of complex, skill-demanding heterogeneous tasks? \textbf{(RQ1)}

    \item How can this balanced workforce composition feature ensure sustainable volunteer retention by valuing their contributions through reimbursement or recognition, while also modeling their latent potential for expected future endeavours? \textbf{(RQ2)}

\begin{figure*}[!t]
\centering
{\includegraphics[width=\textwidth]{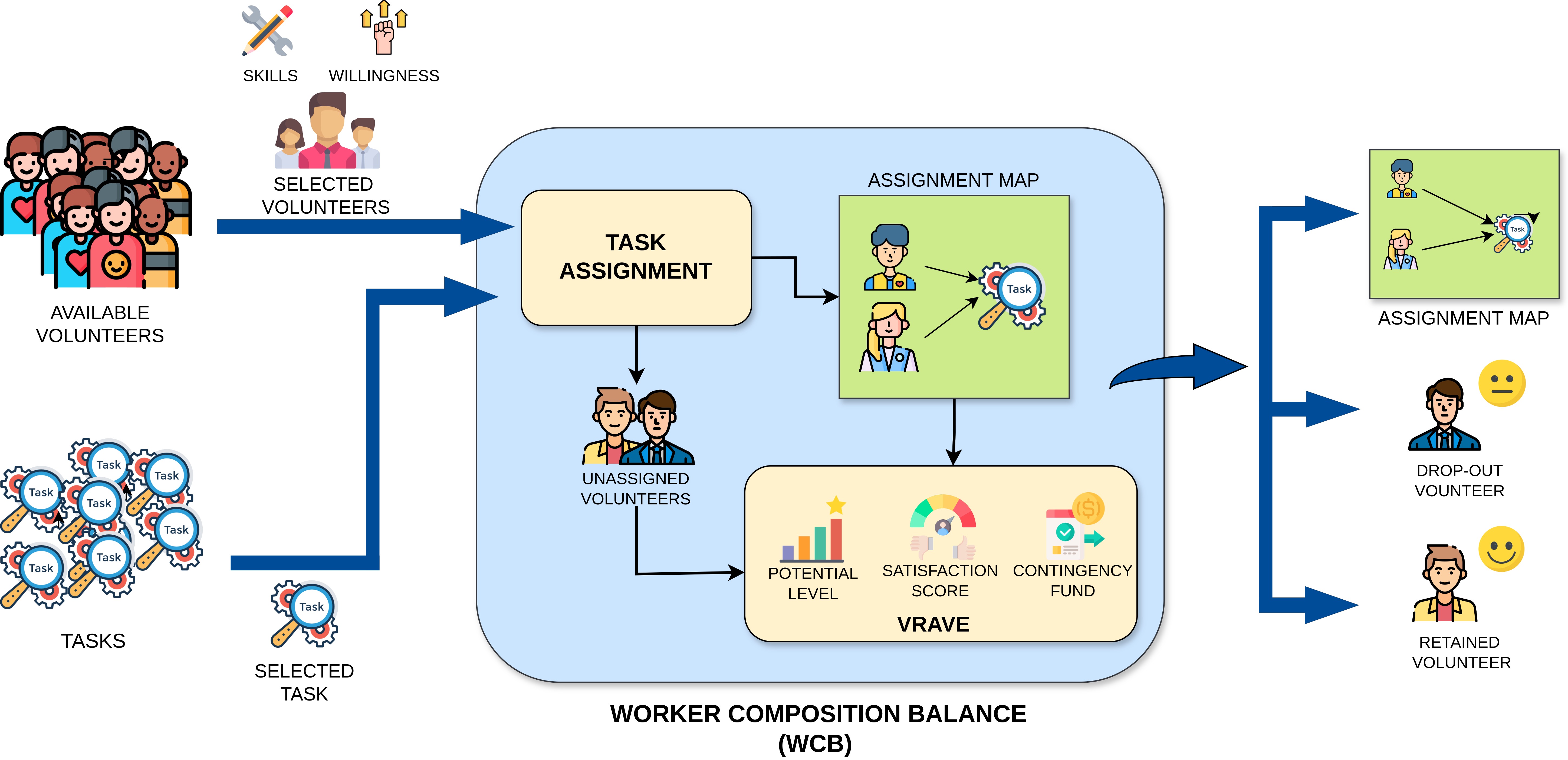}}
\vspace{-0.15in}
\caption{Proposed Worker Composition Balance (WCB) Framework: Integrating Task Assignment Mechanism with Volunteer Retention and Value Enhancement (VRAVE) Module }
\label{frame}
\vspace{-0.2in}
\end{figure*}
\end{itemize}

\vspace{-0.05in}
\subsection{Contributions}
\vspace{-0.05in}
In this paper, the aforementioned research questions are addressed by focusing on developing a workforce composition balance policy framework aimed at retaining potential volunteers for sustained participation. The policy is specifically designed to seamlessly integrate with the existing skill-oriented task assignment strategy. For our methodology, we chose the state-of-the-art \textit{SWill-TAC} \cite{samanta2021swill} as the foundational skill-oriented task assignment method. \textit{SWill-TAC}, a skill-oriented dynamic task allocation approach with a willingness factor for complex assignments, iteratively assigns workers (volunteers) to tasks based on the required skills and their skill sets. \textit{SWill-TAC} also considers the workers' willingness to perform tasks and monitors the budget constraints associated with the tasks. In addition to our proposed retention mechanism, we conducted experiments with three baseline retention mechanisms proposed by the authors in \cite{Difallah2014}. All the experiments are conducted on a real life dataset i.e \textit{UpWork} \cite{upwork}.
The main contributions of this paper are:

\begin{itemize}
    \item We introduce the \textit{Volunteer Retention and Value Enhancement (VRAVE)} algorithm, which aims to maintain workforce composition and retain potential volunteers within a skill-oriented task assignment policy.
    \item We develop a \textit{Workforce Composition Balance (WCB)} framework that successfully integrates \textit{VRAVE} with the state-of-the-art skill-oriented task assignment method for complex, heterogeneous tasks. The illustrative diagram of the \textit{WCB} framework is shown in \textbf{Figure \ref{frame}}.
    \item We perform extensive experiments using real (\textit{UpWork})  dataset. \textit{VRAVE} is evaluated against three baseline retention mechanisms, with results demonstrating the superiority of our approach.
\end{itemize}

The remainder of the paper is organized as follows. The literature review is presented in Section \ref{section2} while the preliminaries needed for \textbf{WCB} framework design model in Section \ref{section3}. The proposed \textbf{WCB} framework along with the working of \textbf{VRAVE} algorithm is described in Section \ref{section4}. Experiments and performance results are presented in Section \ref{section5}. Conclusion and future works are offered in Section \ref{conc}.

\vspace{-0.05in}
\section{Related Work}
\label{section2}
\vspace{-0.05in}
In this section, we discuss existing work related to workforce composition balance and retention policies in crowdsourcing (CS).

The authors in \cite{Pilourdault2017} attempted to model worker motivation by examining its effect on different performance metrics in micro-tasks CS. They allowed workers to select tasks and proposed three mechanisms: task relevance-based, task diversity-based, and combined diversity-payment-based. However, their focus was primarily on motivating participants who are already working on a batch of micro-tasks but are reluctant to contribute their full potential.

Similarly, \cite{Difallah2014} aimed to reduce overall batch execution latency by retaining participants working on a particular batch of HITs, with experiments conducted on Amazon MTurk. This study concentrated on maintaining worker participation for the duration of specific tasks to improve efficiency.

In \cite{Kaur2022}, the authors created an integer program to help non-profit organizations generate volunteer and employee task assignments for specific program planning horizons. This model takes into account volunteer retention needs, as well as organizational and task urgency needs when assigning tasks.

However, to the best of our knowledge, our paper is the first to propose a framework for maintaining workforce composition in CS, focusing on retaining \textit{potential} volunteers through a \textit{skill-oriented task assignment} approach. 

\vspace{-0.05in}
\section{Preliminaries}
\label{section3}
\vspace{-0.05in}
We proceed to present necessary definitions and preliminaries needed for our framework. 

\begin{definition} (\textbf{Task}):
A task, denoted by $\mathbf{t = (B_{t}, S_{t}, \delta_{t}, e_{t})}$, encompasses the following attributes: (i) Budget($B_{t}$), representing the allocated budget for task $t$; (ii) Skill Set ($S_{t}$), denoting the set of skills required to complete task $t$; (iii) Arrival Time ($\delta_{t}$), indicating the time at which task $t$ arrives; and (iv) Expiration Time ($e_{t}$), specifying the time by which task $t$ is expected to be completed, leading to an expiration time of ($\delta_{t} + e_{t}$).
\end{definition}

\begin{definition}(\textbf{Volunteer}):
A volunteer, denoted by $\mathbf{v = (C_{v}, S_{v},\delta_{v}, e_{v})}$, possesses the following attributes: Expense ($C_{v}$), indicating the costs $v$ needs to bear while executing tasks, such as travel allowances, tools, stationery, and bills, which are expected to be covered by the remuneration received from the assigned task's budget; Skill Set ($S_{v}$), representing the set of skills that volunteer $v$ has;  Availability Time ($\delta_{v}$), denoting the time from which $v$ is available; and Departure Time ($e_{v}$), specifying the time at which $v$ is expected to leave or stop being available.
\end{definition}

The proposed framework operates under the assumption that a task's completion depends on the coverage of skills required for that task. However, volunteers differ from regular paid crowd workers in that they receive a remuneration $RW_{v}$ specifically to cover their incurred expense $C_{v}$. Moreover, it is important to note that $RW_{v}$ is not always equal to $C_{v}$. We also assume that $C_{v}$ remains the same for any given task assigned to volunteer $v$. Similar to the approaches taken by authors in \cite{Song2020,samanta2021swill, Sama2212:Volunteer,Cheng2016}, we consider that the skills required for a task or possessed by a volunteer are always part of a preset universal skill set $S$ with a constant size. Additionally, following the concept of one-to-many mapping as discussed in \cite{samanta2021swill}, it is possible for a task to have multiple volunteers at a given time frame, but the reverse scenario is not true.

\begin{definition} (\textbf{Execution Round}) 
The execution round, denoted by \( r \), follows a batch-based operation policy within the simulation environment, where logical time units are divided into intervals \( \{ i_{0}, i_{1}, \ldots, i_{n} \} \). At the commencement of each interval \( r = i_{k} \), the chosen task assignment mechanism executes, taking into account the volunteers and tasks received during the previous round \( (r-1) = i_{k-1} \). Subsequently, our proposed workforce composition balance mechanism is executed, which considers the assignment map generated by the selected task assignment mechanism during the current round. 
\end{definition}

Our proposed \textit{VRAVE} algorithm is designed to be preferably integrated with online task assignment methods that operate in dynamic environments, where the arrival and departure patterns of tasks and volunteers follow specific distributions (see Section \ref{section5} for Datasets Description). To ensure stable and consistent results, the \textit{VRAVE} algorithm is triggered at regular time intervals, marking each execution round denoted as \( r \).

\begin{definition} (\textbf{Potential Level}): 
The Potential Level, denoted by $\Pi(v,r)$, serves as a predictive metric estimating the probability of a volunteer $v$ being a valuable candidate for future assignments within execution round $r$. The $\Pi$ function takes two arguments: $v$ represents the volunteer associated with $\Pi$, and $r$ denotes the specific execution round under consideration. It is important to note that $\Pi(v,r)$ is revisited at the beginning of each new execution round to reflect updated information.

The formula for Potential Level is given by:
\begin{equation}
    \Pi(v,r) = (1 - \Pi(v,r-1)) \cdot \Pi_{init}(v,r) + \Pi(v,r-1)
\end{equation}
\end{definition}
Here, $\Pi_{init}$, an initialization function, plays a crucial role in assigning the initial potential value. As $v$'s $\Pi$ increases, indicating greater experience and efficiency, $v$ becomes more resourceful and valuable for assignments within round $r$. It is worth noting that when newcomers join at round $r-1$, their $\Pi$ at that stage is assumed to be null, emphasizing the significance of $\Pi_{init}$ in determining initial potential levels.

$\Pi_{init}$ is calculated by the following formula:
\begin{equation}
\label{eqpot}
    \Pi_{init}(v,r)=\frac{1}{1+\exp^{(-\sigma_{(v,r)})}} 
\end{equation}
\begin{equation}
\label{eqinit}
    \text{where,} \:\: \sigma_{(v,r)} = \left( \frac{Alloc^{succ}_{v}}{Alloc^{par}_{v}} \right) + \left(\frac{|S_{v} \cap S|}{|S|} \right) + \psi_{v}
    \end{equation}
\begin{equation}
    \text{And}\:\: \psi_{v}=\alpha^{1/l_{v}}
\end{equation}
To begin, $\sigma_{(v,r)}$ comprises three additive components. The first component is the ratio of the number of times a volunteer $v$ has been successfully selected ($Alloc^{succ}_{v}$) to the total number of allocation rounds $v$ has participated in so far ($Alloc^{par}_{v}$). This ratio helps gauge the historical performance of volunteers. The second component is the ratio of the skills possessed by $v$ to the total number of skills in the universal skill set $S$. This ratio reflects the current competencies of volunteers given the diverse task requirements in the pool. Many CS platforms, such as Upwork\cite{upwork}, provide skill development resources through which crowd-workers can update and enhance their skill sets. Therefore, our framework assumes that over execution rounds, the skill set $S_{v}$ of volunteers will expand.

The third component is the \textit{Aging Factor}, denoted as $\psi_{v}$. Generally, volunteers asking for lower remuneration or have higher potential (like experienced) are preferred during allocation, leading to repeated neglect of those requesting higher pay or lower potential (like newcomers). However, the latter volunteers may be more suitable for upcoming tasks in following rounds. This repeated neglect could lead to resultant starvation. To address this issue, we introduce the \textit{Aging Factor} in the $\Pi_{init}$ formulation.

The \textit{Aging Factor} incorporates the \textit{aging constant} $\alpha$, which ranges from $(0,1]$. The input $l$ represents the number of rounds that have passed since a volunteer $v$ was last successfully assigned to a task. As $v$ remains unallocated for more rounds, $\psi_{v}$ increases. For \textit{newcomers}, the \textit{aging constant} is set to $0.1$. The concepts of $\psi$ and $\Pi_{init}$ are inspired by operating system principles and related works \cite{silberschatz2006operating,lindgren2004probabilistic,Doria2015}.

The value of $\sigma_{v,r}$ always falls within the range of $[0,3]$. For any positive value of $\sigma_{v,r}$ less than or equal to three, $\Pi_{init}(v,r)$ should be a \textit{monotonic increasing function} that ranges between $[0,1]$. This relationship ensures that the potential level $\Pi_{level}$ remains within the bounds of 0 to 1.

\begin{definition} (\textbf{Participation Dividend})
The participation dividend for a volunteer $v$, denoted as $D_{v}$, is an allowance rewarded to a volunteer who has not been allocated tasks for two or more consecutive execution rounds. $D_{v}$ is calculated to acknowledge the continued engagement and improvement in the volunteer's potential level $\Pi(v,r)$ over the rounds. The formula for $D_{v}$ is as follows:
\begin{equation}\label{eqdiv}
\begin{split}
   D_{v} = \gamma \cdot \left( \frac{\Pi(v,r-1)}{\max_{\forall v' \in V \text{ at } r} \Pi(v',r-1)} \right) &\cdot \\
   \left( \frac{U}{\sum_{v \in {V_{unassigned}}} \left( \frac{\Pi(v,r-1)}{\max_{\forall v' \in V \text{ at } r} \Pi(v',r-1)} \right)} \right)
\end{split}
\end{equation}

\end{definition}
In this formula, $U$ represents the contingency amount set aside by the CS platform to pay the dividend to volunteers. $\gamma$ is a scaling parameter that adjusts the overall level of the dividend distributed to volunteers. It ensures that the dividend distribution remains within the allocated contingency $U$ while still providing meaningful incentives to volunteers. The calculation of $D_{v}$ ensures that volunteers with higher potential levels receive a higher dividend share, and the total dividend distributed among all unassigned volunteers ($V_{unassigned}$) does not exceed the allocated contingency $U$.

\begin{definition} (\textbf{Satisfaction Score})
Satisfaction score, denoted as $Sat(v,r)$ is a quantitative measure used to assess the level of contentment, fulfillment, or happiness experienced by volunteer $v$ at execution round $r$. The formula for $Sat(v,r)$ is:
\begin{equation}\label{eqsat}
    \text{Sat}(v,r) =(1-\omega).\left( \frac{\Pi(v,r-1)}{\max_{\forall v' \in V \text{ at } r} \Pi(v',r-1)} \right) + \omega.\left( \frac{CI(v,r)}{C_{v} \times (r-1)} \right)
\end{equation}
\end{definition}
In this formula, $\omega$ is a weighting parameter between $[0,1]$ that determines the influence of the potential level and cumulative income on computing the satisfaction score and the resulting $Sat(v,r)$ is always between 0 and 1. The first component of $Sat(v,r)$ is determined by comparing the potential level of volunteer $v$ at round $(r-1)$ against the maximum potential level achieved by any volunteer at round $r$. On the other hand, the second component measures the ratio of $v$'s cumulative income until round $r$, denoted as $CI(v,r)$, to the total payment received by $v$ until round $r$. Considering, that volunteers will receive full cost coverage when assigned task is completed (i.e $RW_{v}=C_{v}$), this cumulative income $CI(v,r)$ is calculated as follows:
\begin{equation}
    CI(v,r)= CI(v,r-1)+ C_{v}+D_{v}
\end{equation}

\vspace{-0.2in}
\section{Workforce Composition Balance Framework}
\label{section4}
\vspace{-0.05in}
In this section, we build upon the preliminary definitions presented earlier to delve into the workflow of our proposed Workforce Composition Balance (WCB) Framework. We initiate by outlining the \textit{Volunteer Retention and Value Enhancement} (VRAVE) algorithm (refer to Algorithm \ref{algo3}), which addresses research question \textbf{RQ2}. Following that, we explore the seamless integration of VRAVE with a skill-oriented task assignment method tailored for managing complex heterogeneous tasks. Specifically, we highlight SWill-TAC \cite{samanta2021swill} as our chosen foundational skill-oriented task assignment method within this framework. Therefore, this integration feature of the WCB framework provides the answer to research question \textbf{RQ1}. It is worth noting that the workflow of our framework can also accommodate other skill-oriented task assignment method in place of SWill-TAC.

The SWill-TAC algorithm operates by evaluating the most suitable candidate (volunteer) for a given task based on their skill set, willingness, and task budget. When considering tasks, the algorithm scans for candidates whose skills align with the task requirements and computes a utility value for each candidate. The candidate with the highest utility value is then chosen for task allocation. Conversely, for candidates, the algorithm searches for tasks matching their skill set and initiates a recursive process for task allocation. Subsequently, an allocation map $M$ is generated, where $M(t)=\{v_{1},v_{2},..v_{n}\}$ $\forall t \in T$ and $\forall v \in V$, with $n \leq |V|$.

Regarding the attributes utilized in SWill-TAC, they correspond directly to the \textit{Task} and \textit{Volunteer} definitions outlined in Section \ref{section3}. The calculation of potential levels and satisfaction scores for volunteers in set $V$ is straightforward. The allocation map $M$ essentially functions as a dictionary that maps tasks to the volunteers assigned to them.

\begin{algorithm}[h]
    \caption{VRAVE (Volunteer Retention and Value Enhancement)}
    \label{algo3}
    \scriptsize
    \begin{algorithmic}[1]
        \renewcommand{\algorithmicrequire}{\textbf{Input:}}
        \renewcommand{\algorithmicensure}{\textbf{Output:}}
        \REQUIRE Contingency $U$, Round $r$, Allocation map $M$, Current volunteer set $V$, Task set $T$
        \ENSURE Drop-out volunteers set $V_{drop}$, Retained volunteers set $V_{retain}$, Total dividend $Div$
        
        \STATE \textit{Initialization}: Initialize $V_{drop} \gets \varnothing$, $V_{retain} \gets \varnothing$, $Div \gets 0$
        \STATE Generate set of current unassigned volunteers $V_{unassigned}$ from $M$ and $V$.
        
        \FOR{each $v \in V_{unassigned}$}
            \STATE Calculate dividend $D_{v}$ using Equation-\ref{eqdiv}
            \STATE $Div \gets Div + D_{v}$
            
            \STATE Calculate satisfaction score $Sat(v,r)$ using Equation-\ref{eqsat}
            \IF{$Sat(v,r) < threshold$}
                \STATE $V_{drop} \gets V_{drop} \bigcup \{v\}$
                \STATE print ``$v$ quits the platform"
            \ELSE
                \STATE $V_{retain} \gets V_{retain} \bigcup \{v\}$
            \ENDIF
        \ENDFOR
        
        \STATE Return $V_{drop}$, $V_{retain}$, $Div$
    \end{algorithmic}
\end{algorithm}

Algorithm \ref{algo3} presents the pseudo-code for VRAVE. For each volunteer in the unassigned volunteer set $V_{unassigned}$, which is derived from scanning $M$ and $V$, the participation dividend is computed using Equation \ref{eqdiv}, and the satisfaction score $Sat(v,r)$ is evaluated using Equation \ref{eqsat}. If $Sat(v,r)$ falls below a predefined $threshold$, the volunteer $v$ is added to the dropout list $V_{drop}$; otherwise, they are included in the retain list $V_{retain}$. Finally, the algorithm returns $V_{drop}$, $V_{retain}$, and the total dividend amount $Div$.

As detailed in Section \ref{section3}, our framework operates in execution round mode. Therefore, at each round $r$, the task assignment mechanism is executed first, followed by the retention phase in a pipeline sequence. The main driver module of the framework is shown as pseudo-code in Algorithm \ref{algo4}. This algorithm outlines the process for each execution round $r$, starting with initializing the necessary variables. The SWill-TAC algorithm is then executed to assign tasks to volunteers, generating an allocation map $M$. Next, the VRAVE algorithm is applied to $M$ to determine retained and dropped volunteers. Finally, the algorithm updates and returns allocation map $M$, retained volunteers $V_{retain}$, and dropped volunteers $V_{drop}$, leftover contingency fund $U$ and total dividend spent $Div$ for round $r$.

\begin{algorithm}[!t]
 \caption{WCB (Workforce Composition Balance ) Driver Module} \label{algo4}
  \scriptsize
 \begin{algorithmic}[1]
 \renewcommand{\algorithmicrequire}{\textbf{Input:}}
 \renewcommand{\algorithmicensure}{\textbf{Output:}}
 \REQUIRE Task set $T$, Volunteer set $V$, Contingency $U$, Threshold $threshold$
 \ENSURE Allocation map $M$, Retained Volunteers $V_{retain}$, Dropped Volunteers $V_{drop}$, Current Contingency $U$, Dividend amount $Div$
 \\
 \textit{Initialization}: Initialize $M \gets \varnothing$, $V_{retain} \gets \varnothing$, $V_{drop} \gets \varnothing$
 \STATE Start Execution Round $r$
 \STATE $M$ = SWill-TAC ($V$, $T$) // Execute Task Assignment phase to generate allocation map $M$
 \STATE $V_{drop}$, $V_{retain}$, $Div$ = VRAVE($U$, $r$, $M$, $V$, $T$)
 \STATE Update $V_{retain}$, $V_{drop}$, and $Div$ based on VRAVE output
 \STATE Update $U = U - Div$
 \STATE End Execution Round $r$
 \STATE Return $M$, $V_{retain}$, $V_{drop}$, $U$, and $Div$
 \end{algorithmic}
\end{algorithm}

\vspace{-0.1in}
\section{Evaluation}
\label{section5}
\vspace{-0.05in}
The proposed model was evaluated through simulations conducted on a Windows 10 operating system, utilising an Intel i3 dual-core CPU operating at 2GHz and 4GB of RAM. The code was written and compiled using the Python environment. 

\subsection{Datasets Description}
\vspace{-0.05in}
We used the \textit{Upwork} dataset mentioned in \cite{samanta2021swill, Sama2212:Volunteer}, following a common practice adopted by authors in \cite{Jiang2020, Fu2015, Yadav2021,samanta2021swill, Sama2212:Volunteer}. UpWork is a marketplace that leverages a trust-based network linking clients and freelancers. The website provides information about available crowd workers (in our case, volunteers) and jobs that have been posted. We used the tags associated with crowd workers' profiles to determine their available skills. The tasks list the required skills, time, and budget. During data pre-processing, we eliminated any duplicate tasks or crowd workers to reduce redundancy. The attributes included in the task category are task name, ID, budget amount, expected completion time in hours, skill specifications, and arrival time. For the crowd worker category, the attributes are crowd worker name, ID, efficiency, skills, demanded payment (will be replicated as incurred expense), and arrival time. The bias factor and efficiency are calculated using a Uniform distribution, \(U(0,1)\) (which would be used by SWill-TAC). Both the task and worker arrival sequences are assumed to follow a Poisson distribution. Since the number of workers is about 15 times larger than the number of tasks, the rate of arrival for workers is estimated at $75$ per unit of time, and the rate for tasks is $5$ per unit of time. The parameter settings used in our experiments are detailed in Table-\ref{table2}.

\begin{table}[h]
\caption{Parameter settings for experiment}
\label{table2}
\vspace{-0.15in}
\begin{center}
\resizebox{!}{1.2cm}{
\begin{tabular}{|l|l|}
\hline
Parameters & Settings \\
\hline
Number of tasks &  (5-97) \\
Number of volunteers & (250-1575) \\
Average Skill count per volunteers & 7\\
Average remuneration  & \$39.9\\
Average skills demanded per task & 10\\
Average budget of a task & \$428\\
Average completion time of task & 7.5 logical time units\\
Volunteer Bias, User rating, & \\
Willingness factor & [0-1]\\
Each Round & 50 logical time units\\
\hline
\end{tabular}
 }
\vspace{-0.1in}
\end{center}
\end{table}

\vspace{-0.05in}
\subsection{Baselines}
\vspace{-0.05in}
To the best of our knowledge, our work is the first to address the maintenance of workforce composition in Crowdsourcing (CS), focusing on retaining potential volunteers (or workers) within a skill-oriented task assignment paradigm. Existing research has explored incentive-based retention policies without considering the potential level status and skill dimensionality of volunteers or workers.

One such state-of-the-art work is presented by Difallah et al. in \cite{Difallah2014}. In their study, they introduce pricing schemes aimed at enhancing the retention rate of workers engaged in long batches of similar tasks. They investigate how varying the monetary reward over time affects the willingness of workers to complete tasks within a batch. Their work compares new pricing schemes against traditional methods, such as offering a constant reward for all tasks in a batch, and empirically demonstrates how certain schemes effectively incentivize workers to continue working on a given batch of tasks for an extended period. Consequently, we developed three baseline models by integrating the assignment logic of SWill-TAC with the incentive policies from \cite{Difallah2014}.\\
(1) \textbf{Fixed Bonus}: In this approach, volunteers are compensated in a consistent manner for each task they complete during the execution phases. The remuneration amount is presumed to be equivalent to the cost incurred by the volunteer ($C_{v}$). \\ 
(2) \textbf{Training Bonus}: This method implements a linearly decreasing pricing scheme, in which the available budget is allocated to tasks. Initially, the rewards are higher, but they progressively decrease as more tasks are completed.\\
(3) \textbf{Increasing Bonus}: In this case, the pricing scheme provides a gradual increase in reward as the phases progress. In contrast to the commencement of the execution, volunteers receive increased compensation as the rounds progress.

Only SWill-TAC's inherent remuneration policy is amended in accordance with \textit{fixed}, \textit{training}, and \textit{increasing} bonus levels for all of the aforementioned baselines. It is crucial to be aware that the three baselines adhere to their respective remuneration strategy, which is to pay only the current successfully assigned volunteers for the given rounds, without considering the individuals who have consistently remained unassigned in spite of potential level and satisfaction score evaluations. We conducted a comparison between all three baselines and SWill-TAC with VRAVE using our WCB framework.

\vspace{-0.05in}
\subsection{Results and Discussion}
\begin{figure}[h]
\subfigure[Average satisfaction score]{\includegraphics[width=3in,height=2.2in]
{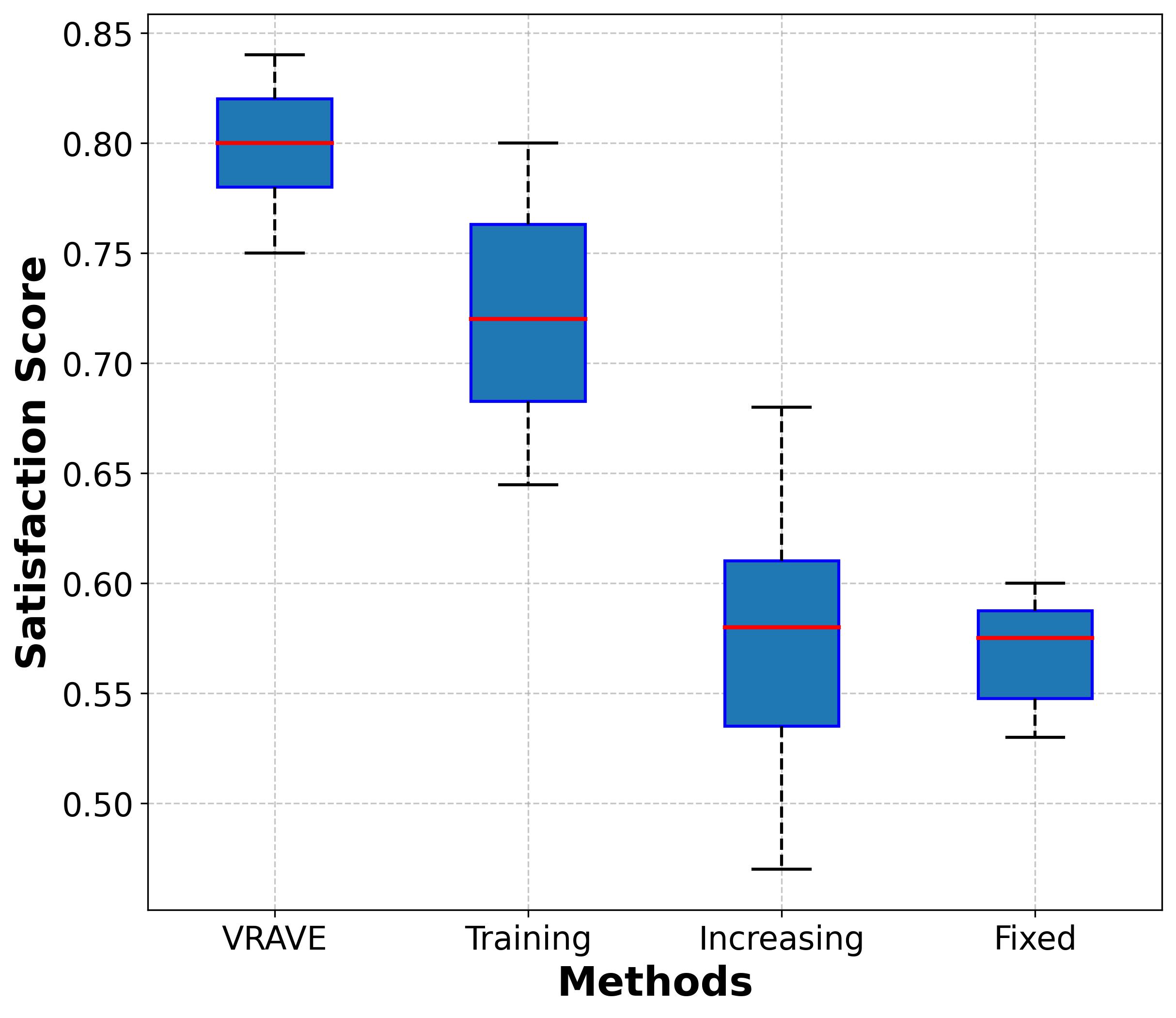}} 
\caption{Average satisfaction scores over six execution rounds. The comparison is between the proposed VRAVE and three baselines, with \textit{SWill-TAC} used as the task assignment mechanism.}
\label{fig2}
\vspace{-0.1in}
\end{figure}

\vspace{-0.05in}
The VRAVE algorithm is dependent on a pre-defined $thresold$ value to determine whether or not to retain a volunteer. In order to determine an appropriate value for this $thresold$, we initially executed VRAVE integrated with SWill-TAC for six consecutive cycles. The satisfaction score of the volunteers was the only measurement collected, and no retention decision was made. Each round was repeated fifty times. The satisfaction scores for the six phases are depicted in \textbf{Figure \ref{fig2}}. We are currently focused solely on VRAVE. We discovered that the median satisfaction score for VRAVE is $0.8$, and the interquartile range (IQR) for VRAVE is $0.039$. Consequently, we established the VRAVE $threshold$ value at $0.75$, which is marginally lower than the median of $0.8$. This guarantees that the $threshold$ captures a significant portion of the data while still allowing for some variability. Ultimately, the $threshold$ value selected may be contingent upon the VRAVE algorithm's specific requirements and objectives, as well as any domain-specific factors.

In addition to the \textbf{satisfaction score}, we employed the metrics of \textbf{retention volunteer count, completed task counts,} and \textbf{average remuneration paid} amount to compare the baseline with WCB (i.e. SWill-TAC with VRAVE). In terms of satisfaction score, VRAVE outperforms these baselines by a significant margin, with an approximate ratio of $1.39$ times higher satisfaction compared to $Training$, $1.4$ times higher compared to $Increasing$, and $1.38$ times higher compared to $Fixed$.  This is due to the fact that satisfaction score is contingent upon both the cumulative incentives gain and the potential level, which marks the coverage of both extrinsic and intrinsic reward policies for satisfaction score assessment. Additionally, the potential level is dynamic and adaptable, taking into account historical factors such as job completion rate, skill development over rounds, and aging (see Eq. \ref{eqpot} and \ref{eqinit}). This definition prioritizes the \textit{experienced} worker while simultaneously nurturing the \textit{newcomers} and preventing the anticipated starvation among them.

\begin{figure*}[h]
\hspace{0.3in}\subfigure[Completed Tasks]{\includegraphics[width=0.45\textwidth,height=2.1in]{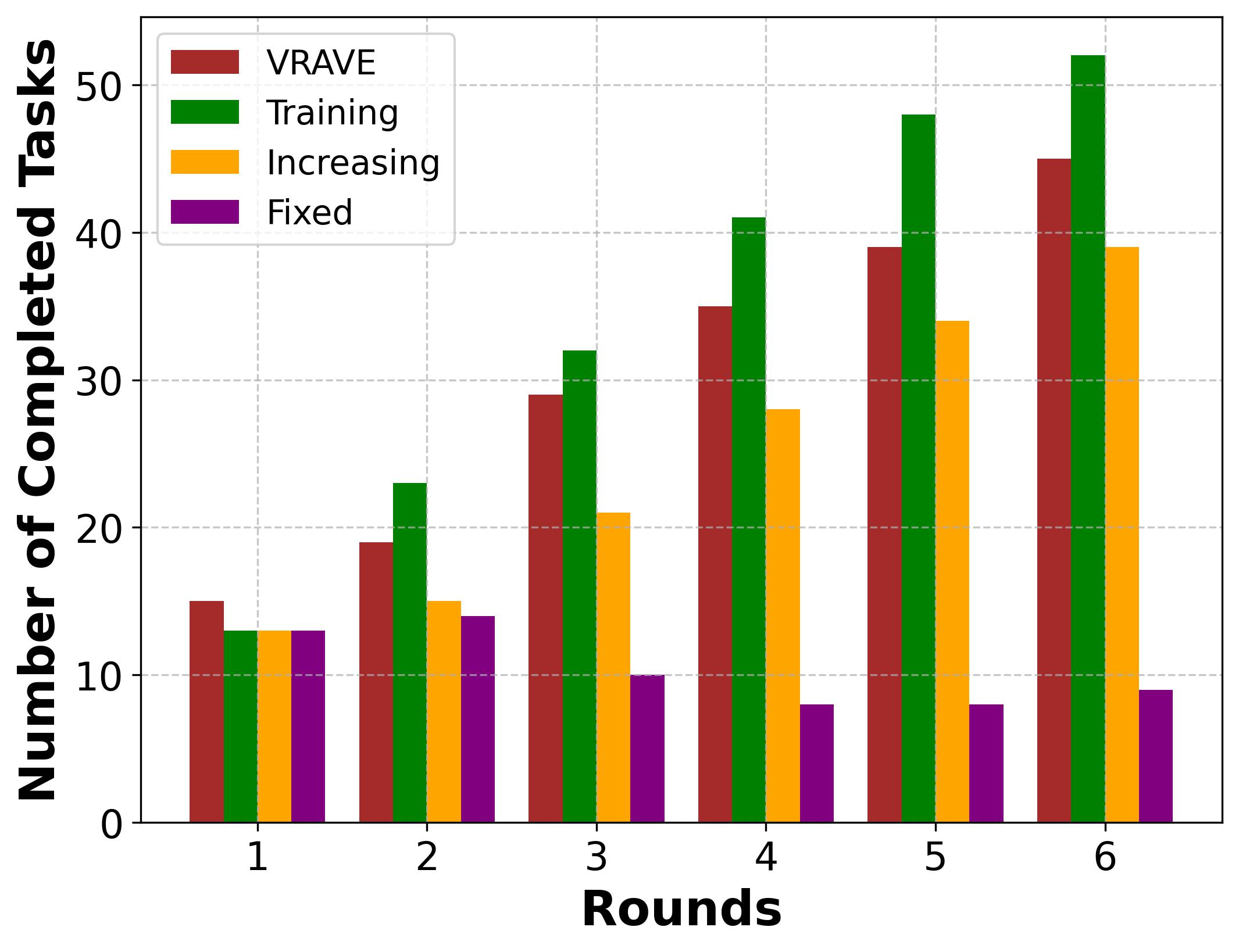}}\label{fig1a}
\subfigure[Volunteer Retention]{\includegraphics[width=0.45\textwidth,height=2.2in]{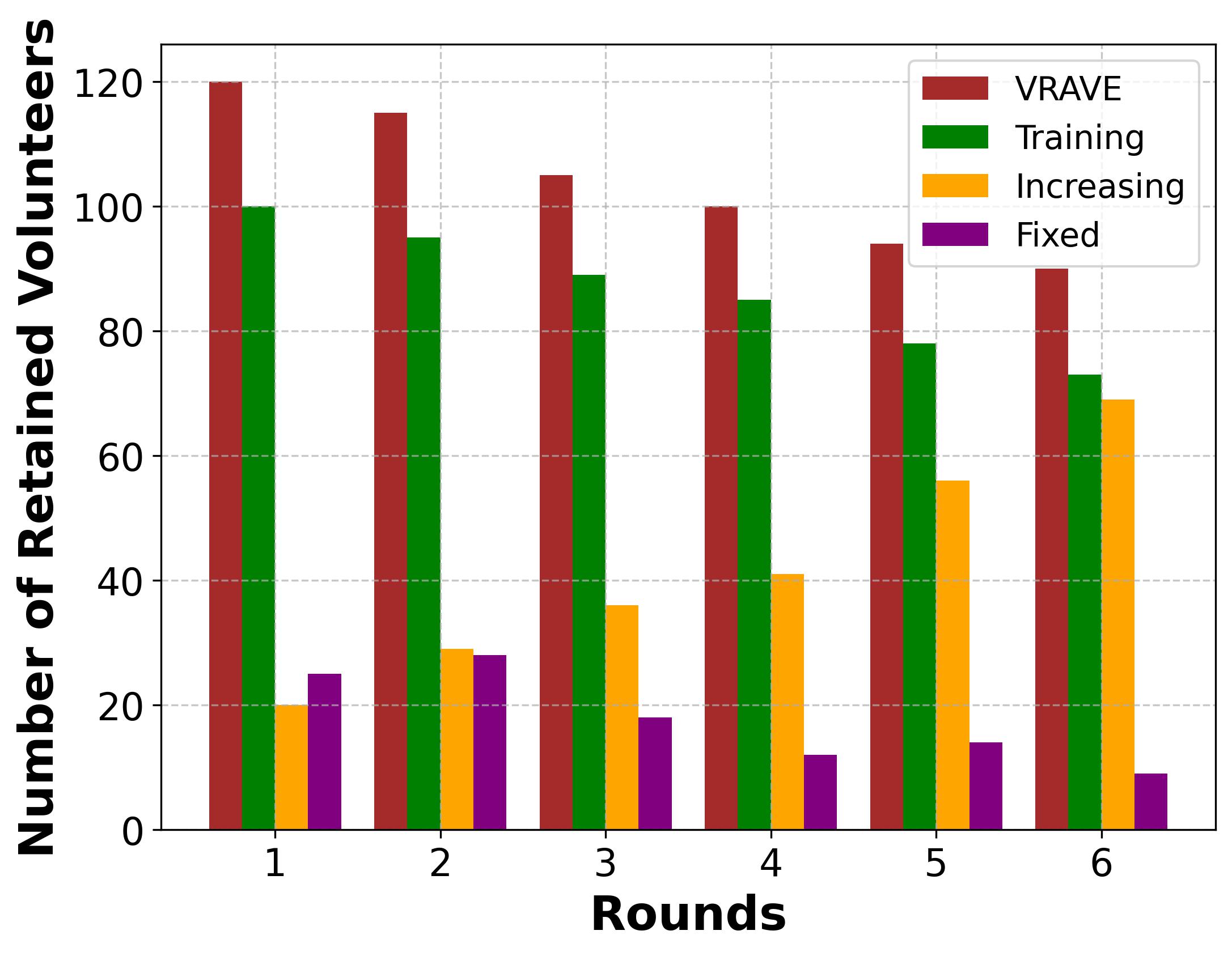}}\label{fig1b} 
\caption{Total completed tasks and  retained volunteers per six execution rounds. The comparison is between the proposed VRAVE and three baselines, with \textit{SWill-TAC} used as the task assignment mechanism.}
\label{fig1}
\vspace{-0.1in}
\end{figure*}

From the \textbf{ Figure \ref{fig1}}, it is observed that VRAVE demonstrates an average task completion rate of $30.5$, while \textit{Training} has an average task completion rate of $32.83$. So, in terms of average task completion, \textit{Training} outperforms VRAVE. However, when comparing VRAVE with \textit{Increasing} $(24) $and \textit{Fixed} $(10.33)$, it shows higher average task completion rates, indicating its superiority over these methods in task completion. Similarly, in terms of volunteer retention, VRAVE's average of $104$ retained volunteers exceeds that of \textit{Training} $(86.67)$, \textit{Increasing} $(42.83)$, and \textit{Fixed} $(18.67)$, demonstrating its effectiveness in retaining volunteers.

\begin{figure}[!t]
    \centering
    \subfigure[Average Remuneration Offered]{\includegraphics[width=3in,height=2.2in]{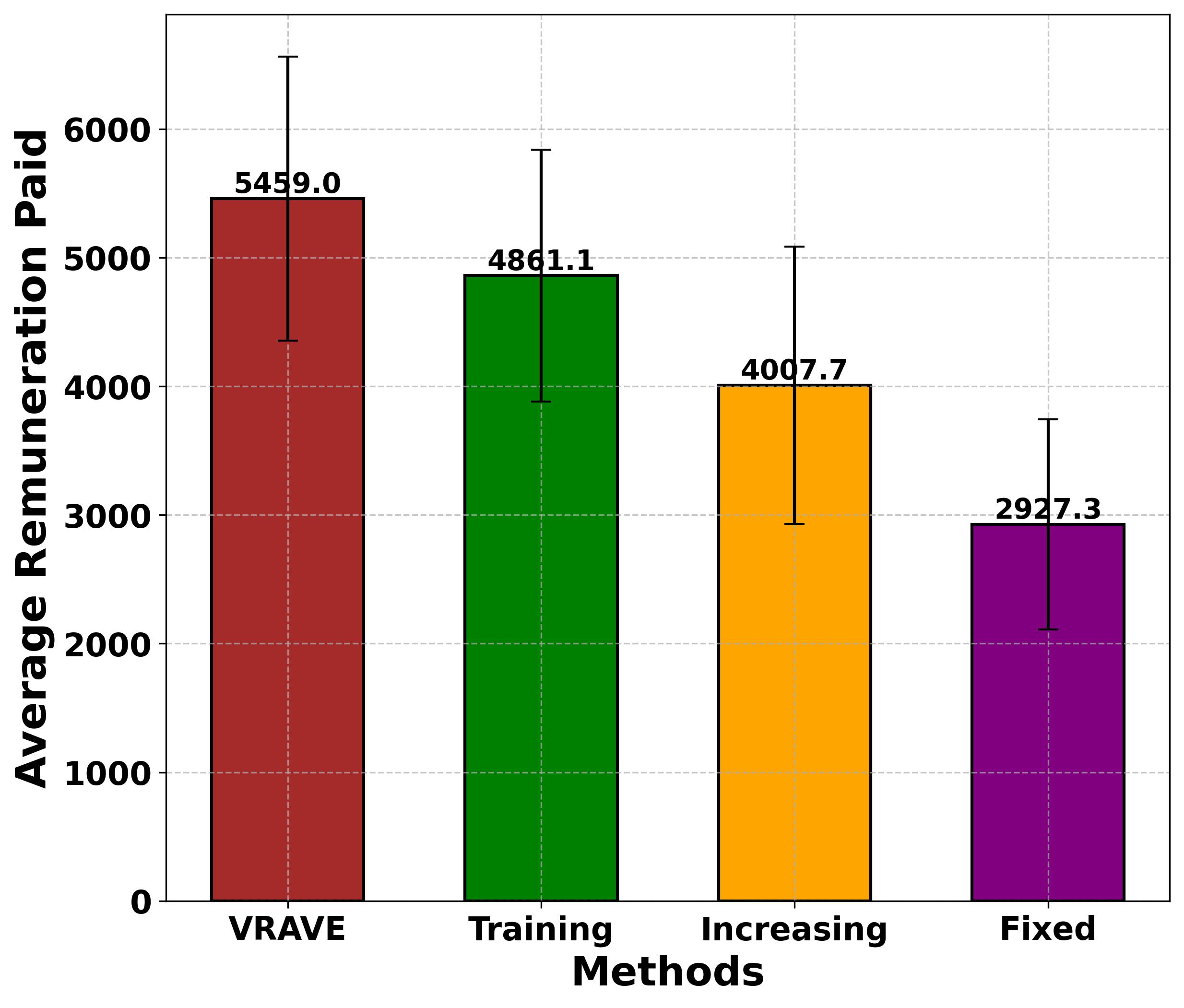}} 
    \caption{Average remuneration paid over six execution rounds. The comparison is between the proposed VRAVE and three baselines, with \textit{SWill-TAC} used as the task assignment mechanism.}
    \label{fig3}
    \vspace{-0.05in}
\end{figure}

Upon analyzing the results based on the average remuneration paid to the volunteers from \textbf{Figure \ref{fig3}}, it is evident that VRAVE has incurred the highest expenditure compared to the baselines. This is attributed to VRAVE's remuneration policy, which compensates both successfully assigned volunteers and those who remain unassigned for two or more consecutive rounds. In contrast, the baselines only compensate the allocated volunteers per round. VRAVE's remuneration mechanism entails direct costs incurred by the assigned volunteers, following a similar policy to \textit{SWill-TAC} assignment.

Additionally, a separate \textit{participation dividend} (see Definition 5) is rewarded to the unassigned volunteers. However, owing to VRAVE's efficient participant dividend policy, the overall remuneration expense is only $ 12\%$ more than that of \textit{Training} (second-best performer with respect to other metrics). During experimentation, for simplicity, it was assumed that the contingency amount $U$ in our \textit{WCB} framework is derived from the leftover budget of all completed tasks, i.e., $U = U + \sum_{t}(B_{t}-RW_{t})$ for all tasks in allocation map $M$ up to the current round $r$. 

\vspace{-0.05in}

\section{Conclusion and Future Work}
\label{conc}
\vspace{-0.05in}
Task assignment in crowdsourcing (CS) has been extensively studied to ensure the right candidates are matched with appropriate tasks. While initial participation is vital, sustained crowd engagement is crucial for the long-term success of CS platforms and the continuous delivery of high-quality services for social good. Existing research has explored incentive-based retention policies, often overlooking the potential level status and skill dimensionality of crowd workers. Within the Volunteer Crowdsourcing Service (VCS) paradigm, where volunteers face complex, skill-demanding tasks, designing a task assignment and retention strategy is essential for maintaining a sustainable volunteer workforce.

To the best of our knowledge, this paper is the first to propose a framework for maintaining workforce composition in CS, focusing on retaining \textit{potential} volunteers through a \textit{skill-oriented task assignment} approach. By integrating our Volunteer Retention and Value Enhancement (VRAVE) algorithm with a skill-oriented task assignment method in our proposed Workforce Composition Balance (WCB) framework, we offer a robust solution for sustaining volunteer engagement and ensuring the effective completion of diverse and complex tasks. WCB outperforms state-of-the-art methods due to its adaptive nature, dynamically adjusting retention decisions by incorporating factors such as volunteer potential levels, participation dividends, and satisfaction scores. Thus, the proposed WCB approach enhances the volunteer engagement and their long-term retention, thus making it suitable for functioning of social good applications where a potential and skilled volunteer workforce is crucial for sustainable community services.

For future work, we plan to test our WCB framework on diverse datasets and explore the stability analysis of task assignment and retention methods.

\bibliographystyle{IEEEtran}
\bibliography{sample}

\begin{thebibliography}{10}
\providecommand{\url}[1]{#1}
\csname url@samestyle\endcsname
\providecommand{\newblock}{\relax}
\providecommand{\bibinfo}[2]{#2}
\providecommand{\BIBentrySTDinterwordspacing}{\spaceskip=0pt\relax}
\providecommand{\BIBentryALTinterwordstretchfactor}{4}
\providecommand{\BIBentryALTinterwordspacing}{\spaceskip=\fontdimen2\font plus
\BIBentryALTinterwordstretchfactor\fontdimen3\font minus \fontdimen4\font\relax}
\providecommand{\BIBforeignlanguage}[2]{{%
\expandafter\ifx\csname l@#1\endcsname\relax
\typeout{** WARNING: IEEEtran.bst: No hyphenation pattern has been}%
\typeout{** loaded for the language `#1'. Using the pattern for}%
\typeout{** the default language instead.}%
\else
\language=\csname l@#1\endcsname
\fi
#2}}
\providecommand{\BIBdecl}{\relax}
\BIBdecl

\bibitem{bedogni2023towards}
L.~Bedogni, M.~Buferli, and D.~Marchi, ``Towards user behavior forecasting in mobile crowdsensing applications,'' in \emph{Proceedings of the 2023 ACM Conference on Information Technology for Social Good}, 2023, pp. 141--148.

\bibitem{shi2022motivates}
X.~Shi, R.~D. Evans, and W.~Shan, ``What motivates solvers’ participation in crowdsourcing platforms in china? a motivational--cognitive model,'' \emph{IEEE Transactions on Engineering Management}, 2022.

\bibitem{nawaz2019gig}
Z.~Nawaz, J.~Zhang, R.~Mansoor, and A.~Ilmudeen, ``Gig workers in sharing economy: Conceptualizing freelancer value proposition (fvp) in e-lancing platforms,'' \emph{Advances in Management and Applied Economics}, vol.~9, no.~6, pp. 51--75, 2019.

\bibitem{leung2021crowd}
G.~S.-K. Leung, V.~Cho, and C.~Wu, ``Crowd workers' continued participation intention in crowdsourcing platforms: An empirical study in compensation-based micro-task crowdsourcing,'' \emph{Journal of Global Information Management (JGIM)}, vol.~29, no.~6, pp. 1--28, 2021.

\bibitem{wu2022understanding}
W.~Wu, Q.~Yang, X.~Gong, and R.~M. Davison, ``Understanding sustained participation in crowdsourcing platforms: the role of autonomy, temporal value, and hedonic value,'' \emph{Information Technology \& People}, 2022.

\bibitem{volmatch}
\BIBentryALTinterwordspacing
{VolunteerMatch}, 2021, accessed: 2021-03-26. [Online]. Available: \url{https://www.volunteermatch.org/}
\BIBentrySTDinterwordspacing

\bibitem{skillimpact}
\BIBentryALTinterwordspacing
{Skilled Impact}, 2021, accessed: 2021-09-06. [Online]. Available: \url{http://skilledimpact.com/skilled-volunteering/}
\BIBentrySTDinterwordspacing

\bibitem{sultan2022addressing}
A.~Sultan, A.~Segal, G.~Shani, and Y.~Gal, ``Addressing popularity bias in citizen science,'' in \emph{Proceedings of the 2022 ACM Conference on Information Technology for Social Good}, 2022, pp. 17--23.

\bibitem{Chen2019}
X.~Chen, ``{A stable task assignment scheme in crowdsourcing},'' \emph{Proceedings - 22nd IEEE International Conference on Computational Science and Engineering and 17th IEEE International Conference on Embedded and Ubiquitous Computing, CSE/EUC 2019}, pp. 489--494, 2019.

\bibitem{Kaur2022}
\BIBentryALTinterwordspacing
M.~P. Kaur, S.~Smith, J.~A. Pazour, and A.~{Duque Schumacher}, ``{Optimization of volunteer task assignments to improve volunteer retention and nonprofit organizational performance},'' \emph{Socio-Economic Planning Sciences}, no. July 2020, p. 101392, 2022. [Online]. Available: \url{https://doi.org/10.1016/j.seps.2022.101392}
\BIBentrySTDinterwordspacing

\bibitem{samanta2021swill}
R.~Samanta, S.~K. Ghosh, and S.~K. Das, ``Swill-tac: Skill-oriented dynamic task allocation with willingness for complex job in crowdsourcing,'' in \emph{2021 IEEE Global Communications Conference (GLOBECOM)}, 2021, pp. 1--6.

\bibitem{Cheng2016}
P.~Cheng, X.~Lian, L.~Chen, J.~Han, and J.~Zhao, ``{Task assignment on multi-skill oriented spatial crowdsourcing},'' \emph{IEEE Transactions on Knowledge and Data Engineering}, vol.~28, no.~8, pp. 2201--2215, 2016.

\bibitem{Song2020}
T.~Song, K.~Xu, J.~Li, Y.~Li, and Y.~Tong, ``{Multi-skill aware task assignment in real-time spatial crowdsourcing},'' \emph{GeoInformatica}, vol.~24, no.~1, pp. 153--173, 2020.

\bibitem{Hettiachchi2022}
D.~Hettiachchi, V.~Kostakos, and J.~Goncalves, ``{A Survey on Task Assignment in Crowdsourcing},'' \emph{ACM Computing Surveys}, vol.~55, no.~3, 2022.

\bibitem{Sama2212:Volunteer}
R.~Samanta, V.~Saxena, S.~Ghosh, and S.~K. Das, ``Volunteer selection in collaborative crowdsourcing with adaptive common working time slots,'' in \emph{2022 IEEE Global Communications Conference: Selected Areas in Communications: Social Networks (Globecom2022 SAC SN)}, Rio de Janeiro, Brazil, Dec. 2022.

\bibitem{8423113}
W.~Wang, Z.~He, P.~Shi, W.~Wu, Y.~Jiang, B.~An, Z.~Hao, and B.~Chen, ``Strategic social team crowdsourcing: Forming a team of truthful workers for crowdsourcing in social networks,'' \emph{IEEE Transactions on Mobile Computing}, vol.~18, no.~6, pp. 1419--1432, 2019.

\bibitem{Difallah2014}
D.~E. Difallah, M.~Catasta, G.~Demartini, P.~Cudr, and P.~Cudr{\'{e}}-Mauroux, ``{Scaling-up the Crowd: Micro-Task Pricing Schemes for Worker Retention and Latency Improvement},'' \emph{Second AAAI Conference on Human Computation and Crowdsourcing}, no. Hcomp, pp. 50--58, 2014.

\bibitem{upwork}
\BIBentryALTinterwordspacing
{Upwork}, 2021, accessed: 2021-03-26. [Online]. Available: \url{https://www.upwork.com/}
\BIBentrySTDinterwordspacing

\bibitem{Pilourdault2017}
J.~Pilourdault, S.~Amer-yahia, D.~Lee, S.~Roy, J.~Pilourdault, S.~Amer-yahia, D.~Lee, S.~R. M.-a.~T. Assign, J.~Pilourdault, S.~Amer-yahia, and D.~Lee, ``{Motivation-Aware Task Assignment in Crowdsourcing To cite this version : HAL Id : hal-01498801 Motivation-Aware Task Assignment in Crowdsourcing},'' 2017.

\bibitem{silberschatz2006operating}
A.~Silberschatz, P.~B. Galvin, and G.~Gagne, \emph{Operating system principles}.\hskip 1em plus 0.5em minus 0.4em\relax John Wiley \& Sons, 2006.

\bibitem{lindgren2004probabilistic}
A.~Lindgren, A.~Doria, and O.~Schelen, ``Probabilistic routing in intermittently connected networks,'' in \emph{International Workshop on Service Assurance with Partial and Intermittent Resources}.\hskip 1em plus 0.5em minus 0.4em\relax Springer, 2004, pp. 239--254.

\bibitem{Doria2015}
A.~Doria and A.~Lindgren, ``{Poster : Probabilistic Routing in Intermittently Connected Networks MobiHoc Poster : Probabilistic Routing in Intermittently Connected Networks},'' vol.~7, no. October, pp. 1--3, 2015.

\bibitem{Jiang2020}
\BIBentryALTinterwordspacing
J.~Jiang, Y.~Zhou, Y.~Jiang, Z.~Bu, and J.~Cao, ``{Batch allocation for decomposition-based complex task crowdsourcing e-markets in social networks},'' \emph{Knowledge-Based Systems}, vol. 194, p. 105522, 2020. [Online]. Available: \url{https://doi.org/10.1016/j.knosys.2020.105522}
\BIBentrySTDinterwordspacing

\bibitem{Fu2015}
Y.~Fu, H.~Chen, and F.~Song, ``{STWM: A solution to self-adaptive task-worker matching in software crowdsourcing},'' \emph{Lecture Notes in Computer Science (including subseries Lecture Notes in Artificial Intelligence and Lecture Notes in Bioinformatics)}, vol. 9528, pp. 383--398, 2015.

\bibitem{Yadav2021}
A.~Yadav, J.~Chandra, and A.~S. Sairam, ``{A Budget and Deadline Aware Task Assignment Scheme for Crowdsourcing Environment},'' \emph{IEEE Transactions on Emerging Topics in Computing}, vol. 6750, no.~c, pp. 1--14, 2021.

\end{thebibliography}

\end{document}